\documentclass[aps,prb,groupedaddress,twocolumn]{revtex4}

\usepackage{graphicx}
\usepackage{amsmath}
\usepackage{amsfonts}
\usepackage{amssymb}
\usepackage{subfigure}
\usepackage{color}

\begin{document}

\title{Biexciton state preparation in a quantum dot via adiabatic rapid passage: \\
       comparison between two control protocols and impact of phonon-induced dephasing}

 \author{M. Gl\"assl$^{1}$}
 \email[]{martin.glaessl@uni-bayreuth.de}
 \author{A. Barth$^{1}$}
 \author{K. Gawarecki$^{2,3}$}
 \author{P. Machnikowski$^{2}$}
 \author{M. D. Croitoru$^{1}$} 
 \author{S. L\"uker$^{3}$}
 \author{D. E. Reiter$^{3}$}
 \author{T. Kuhn$^{3}$}
 \author{V. M. Axt$^{1}$}

\affiliation{$^{1}$Institut f\"{u}r Theoretische Physik III, Universit\"{a}t Bayreuth, 95440 Bayreuth, Germany}
\affiliation{$^{2}$Institute of Physics, Wroc{\l}aw University of Technology, 50-370 Wroc{\l}aw, Poland}
\affiliation{$^{3}$Institut f\"ur Festk\"orpertheorie, Universit\"at M\"unster, 48149 M\"unster, Germany}

\date{\today}

\begin{abstract}
We investigate theoretically under which conditions a stable and high-fidelity preparation
of the biexciton state in a quantum dot can be realized by means of adiabatic rapid passage
in the presence of acoustic phonon coupling.
Our analysis is based on a numerically complete real-time path integral approach
and comprises different schemes of optical driving using frequency-swept (chirped)
pulses. We show that depending on the size of the biexciton binding energy,
resonant two-photon excitations or two-color schemes can be favorable. 
It is demonstrated that the carrier-phonon interaction strongly affects the
efficiency of the protocols and that a robust preparation of the biexciton is
restricted to positive chirps and low temperatures. A considerable increase
of the biexciton yield can be achieved realizing temperatures below $4$~K. 
\end{abstract}

\maketitle


\section{INTRODUCTION}

Realizing an on demand source of entangled photon pairs is crucial for many
innovative applications in quantum information science \cite{bouwmeester:00}
and quantum optics \cite{haroche:06}. Related applications comprise quantum
computation \cite{nielsen:00}, quantum teleportation \cite{bouwmeester:97},
or quantum key distribution \cite{gisin:02} as well as tests of fundamental
aspects of quantum mechanics. A very promising scheme to generate polarization entangled
photon pairs is to use the radiative decay of the biexciton state (consisting of two electron-hole
pairs) in a semiconductor quantum dot (QD) \cite{moreau:01,stevenson:06}. Problems 
related to the fine structure splitting of the intermediate exciton states have largely
been overcome \cite{akopian:06,trotta:12} and recently, impressive experiments reported the
realization of ultrabright sources by coupling a photonic molecule to a single QD
\cite{dousse:10}.

A necessary precondition for an efficient 
use of the biexciton decay cascade is a robust preparation of the biexciton state.  
In principle, an inversion of the QD from the ground to the biexciton state is possible by
driving the system with a transform limited laser pulse of constant frequency, that is
resonant to half of the ground state biexciton transition frequency \cite{stufler:06,machnikowski:08}.
The drawback of this Rabi-flopping scheme is that it requires a very detailed
knowledge of the system parameters like the transition dipole moments as well as
a very precise control of the field intensity in order to ensure the desired inversion.
Recalling that these requirements are often only insufficiently fulfilled,
it seems that a more practical way to realize a stable and high-quality preparation
of the biexciton state is the use of optical driving schemes that rely on adiabatic
rapid passage (ARP) and use frequency-swept (chirped) pulses.
The basic idea underlying such schemes, that are stable with respect to intensity
changes of the laser field, is to drive the system adiabatically along an 
eigenstate of the coupled light-matter Hamiltonian, whereby the character 
of the state changes during the pulse via a level anticrossing. Recently, experimental 
studies by Wu et al.\cite{wu:11} and Simon et al.\cite{simon:11} that aimed to prepare
the single exciton state  by means of ARP, have proven the applicability of an earlier proposed theoretical 
protocol \cite{schmidgall:10} and demonstrated a stable generation of the exciton 
state, that in the ARP-regime did only slightly depend on the applied pulse area. However, 
the efficiency reported in these experiments stayed below the ideal case 
and most recent theoretical calculations gave compelling evidence that this reduction can be 
attributed to the coupling of the QD to acoustic phonons \cite{lueker:12}. 
It has been demonstrated that acoustic phonons lead to a drastic deterioration of 
the exciton generation for negative values of the chirp, while an efficient 
preparation can be achieved for positive chirps, where the phonon influence is
restricted to phonon absorption processes that become unlikely at low temperatures.\cite{lueker:12} 
The reported drastic changes compared to the ideal scheme in the absence of the carrier-phonon
interaction \cite{schmidgall:10} emphasize the necessity to study the influence of
the unavoidable coupling of carriers to acoustic phonons on other schemes relying on ARP
in order to fully characterize their true potential. 

In this paper, we discuss two different ARP based protocols aiming at an inversion of the
QD to the biexciton state and give a comprehensive analysis of the phonon impact on these schemes.  
In the first protocol, the QD is driven by a single linearly
polarized chirped pulse where the frequency at the pulse maximum
is resonant to half of the ground state biexciton transition
as first suggested by Hui and Liu \cite{hui:08}. In the second scheme, the QD is driven
by two circularly polarized chirped pulses, that at the pulse maxima 
are resonant to the ground state exciton
and to the exciton biexciton transition, respectively. For both protocols, the
evolution of the system along an adiabatic branch is more complex
than for two-level schemes. For example, in contrast to the two-level case, 
ARP is here sensitive to the sign of the chirp even without taking the carrier-phonon
interaction into account. Therefore, it is a priori unclear to what extent the
efficiency is affected by the coupling to acoustic phonons.
It turns out, however, that similar to the results reported in
Ref.~\onlinecite{lueker:12} for the inversion to the exciton state, an efficient preparation
of the biexciton is restricted to positive chirps and low temperatures.
Quite significant differences between both investigated protocols arise for moderate and
large values of the biexciton binding energy $\Delta$, as the pulse area threshold that
has to be exceeded in order to ensure a purely adiabatic evolution rises strongly
with $\Delta$ for the first scheme, whereas it is independent of $\Delta$ for the 
second protocol. Above the respective ARP-thresholds, the biexciton occupation
depends slightly on the pulse area in a nonmonotonic way.

The paper is organized as follows. In Sec.~\ref{theory} we outline the model
and comment briefly on the real-time path integral method that is used for
our numerical simulations. The latter are presented in Sec.~\ref{results}, where 
Sec.~\ref{results:GBT} analyzes the phonon impact on the two-photon
resonance scheme, while Sec.~\ref{results:2PP} deals with the phonon 
influence on the two-color protocol and compares the efficiency of both schemes.
Finally, Sec.~\ref{conclusions} concludes the paper.


\section{THEORY}
\label{theory}

We consider an optically driven strongly confined GaAs QD coupled
to a continuum of acoustic phonons. Our model Hamiltonian is
defined by $H= H_{\rm{dot,las}}^{\rm{circ/lin}}+H_{\rm{dot,ph}}$,
where the first term describes the electronic structure of the
QD and the coupling to a laser of either circular or linear
polarization, while the second term represents the carrier-phonon coupling.
For excitations involving circularly $\sigma_+$ as well as circularly
$\sigma_-$ polarized pulses, we account for four bright states:
the ground state $|G\rangle$, the two single exciton states 
$|\sigma_{\pm}\rangle $, and the biexciton state $|B\rangle$.
In this basis, $H_{\rm{dot,las}}^{\rm{circ}}$ reads
\begin{align}
 \label{eq:Hcirc}
 H_{\rm{dot,las}}^{\rm{circ}}
 &= \hbar\omega_0(|\sigma_{+}\rangle \langle\sigma_{+}|\!+\!|\sigma_{-}\rangle \langle\sigma_{-}|) \\ \notag
 &+ (2\hbar\omega_0\!-\!\Delta)|B\rangle \langle B| \\ \notag
 &+ \frac{\hbar}{2}[f^{\sigma_+}(t) \, (|\sigma_{+}\rangle\langle G| + |B\rangle \langle \sigma_{-}|) + \rm{h.c.}] \\ \notag
 &+ \frac{\hbar}{2}[f^{\sigma_-}(t) \, (|\sigma_{-}\rangle\langle G| + |B\rangle \langle \sigma_{+}|) + \rm{h.c.}] \, ,
\end{align} 
where $f^{\sigma_{\pm}}(t)=2{\bf{M\cdot E}}^{(+)}_{\sigma_{\pm}}(t)/\hbar$ with
${\bf{E}}^{(+)}_{\sigma_{\pm}}$ being the positive frequency component of the $\sigma_{\pm}$
circularly polarized light field and ${\bf{M}}$ denoting the transition dipole element.
$\hbar\omega_0$ defines the ground state exciton transition energy and $\Delta$ represents the biexciton
binding energy. Electron-hole exchange interactions, that result in a direct coupling
between the single exciton states are neglected, which is well justified
for the picosecond time-scale considered here and a sufficiently small fine structure splitting
(within $10\,\mu\rm{eV}$ for the strongest chirps studied here). 
For linearly polarized excitations, it is advantageous to introduce the single
exciton state $|X\rangle = (|\sigma_{+}\rangle + |\sigma_{-}\rangle)/\sqrt{2}$
which allows us to apply a model with only three levels. \cite{machnikowski:08, glaessl:12} 
$H_{\rm{dot,las}}^{\rm{lin}}$ is then given by
\begin{align}
 \label{eq:Hlin}
 H_{\rm{dot,las}}^{\rm{lin}}
 &= \hbar\omega_0 |X\rangle \langle X| + (2\hbar\omega_0\!-\!\Delta)|B\rangle \langle B| \\ \notag
 &+ [\hbar f^{\rm{lin}}(t)/2 \, (|X\rangle\langle G| + |B\rangle \langle X|) + \rm{h.c.}]  \, 
\end{align} 
with $f^{\rm{lin}}(t)=2{\bf{M\cdot E}}^{(+)}_{\rm{lin}}(t)/\hbar$ and
${\bf{E}}^{(+)}_{\rm{lin}}$ denoting the positive frequency component of the
linearly polarized electric field.  
\begin{figure*}[ttt]
 \includegraphics[width=17.cm]{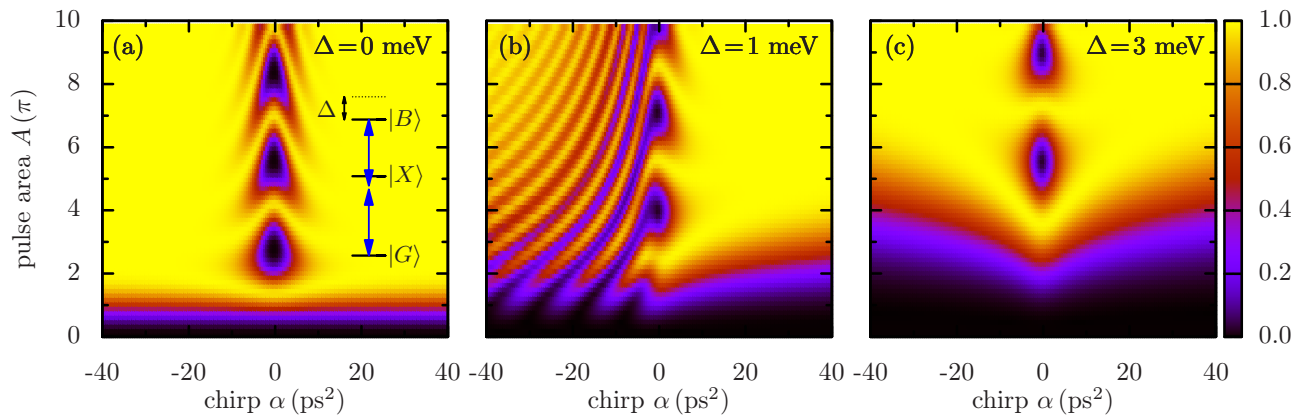}
    \caption{
             (Color online) Final biexciton occupation after a 
             linearly polarized and frequency-swept Gaussian pulse with $\tau_0=2\,\rm{ps}$
             as a function of the original pulse area $A$ and the chirp $\alpha$
             for biexciton binding energies of (a) $\Delta = 0$, (b) $1$, and
             (c) $3\,\rm{meV}$ in the absence of the carrier-phonon coupling. The central frequency
             (at zero chirp) is chosen such that the two-photon process is 
             resonant to the ground state biexciton transition [cf. inset of panel (a)].
            }
 \label{fig1}
\end{figure*}
\begin{figure}[ttt]
 \includegraphics[width=8.5cm]{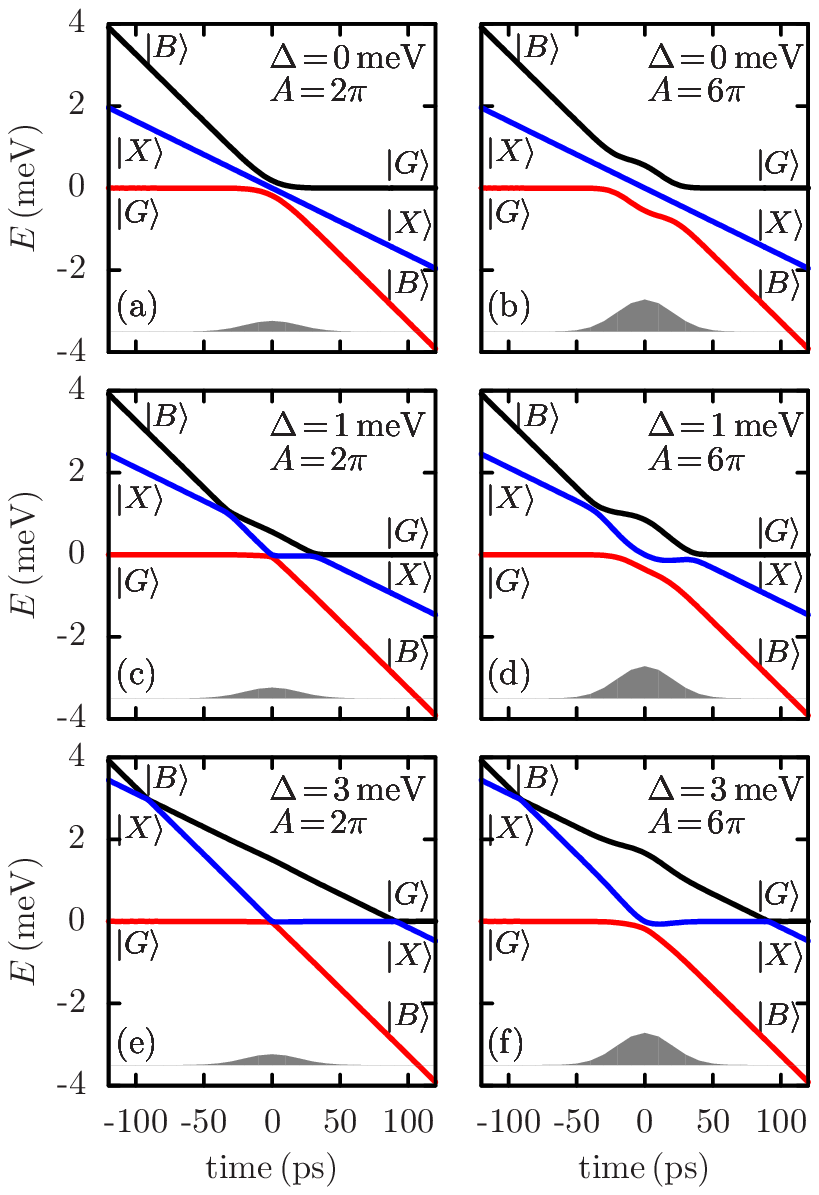}
    \caption{
             (Color online) Adiabatic spectral branches of the laser driven
             QD system for the two-photon resonance ARP-scheme for
             $\tau_0 = 2 \,\rm{ps}$, $\alpha = 40 \, \rm{ps}^2$ and pulse
             areas and biexciton binding energies as indicated. The evolution
             for $\alpha = -40 \, \rm{ps}^2$ can be read off the plots by following
             the branches from the right to the left. The gray shaded areas represent
             the pulse envelope functions, that have been shifted downwards for
             clarity.
            }
 \label{fig2}
\end{figure}

In this work, we will concentrate on optical driving by chirped Gaussian pulses.
These frequency swept pulses can be obtained
from transform limited Gaussian pulses of the form
\begin{align}
 f_0(t) = \frac{A}{\sqrt{2\pi \tau_0^2}} \exp(-\frac{t^2}{2\tau_0^2})\exp(-i\omega t)\, ,
\label{eq:f0}
\end{align}
with the original pulse area $A$, the pulse duration $\tau_0$, and 
the central frequency $\omega$. Here, the polarization label used
in Eqs.~(\ref{eq:Hcirc}) and (\ref{eq:Hlin}) has been suppressed
for simplicity. Applying a Gaussian chirp filter \cite{saleh:07}
with the chirp coefficient $\alpha$ transforms $f_0(t)$ into
\begin{align}
 f(t) = \frac{A}{\sqrt{2\pi\tau_0\tau}} \exp(-\frac{t^2}{2\tau^2})\exp(-i\omega t-i\frac{at^2}{2})\, ,
\label{eq:f}
\end{align}
where $\tau=\sqrt{\alpha^2/\tau_0^2+\tau_0^2}$ characterizes the chirped pulse length
and $a=\alpha/(\alpha^2+\tau_0^4)$ is the frequency chirp rate. 
In the following, we will use pulses of different polarizations
and central frequencies. Details will be given in Sec.~\ref{results}.  

The carrier-phonon coupling is treated in terms of the independent
Boson model \cite{mahan:90}. Concentrating on the deformation potential
coupling to longitudinal acoustic (LA) phonons, which has been shown 
to provide the dominant dephasing mechanism in GaAs QDs
\cite{krummheuer:02, vagov:04, ramsay:10}, $H_{\rm{dot,ph}}$ reads
\begin{align}
\label{Hph}
H_{\rm{dot,ph}} \!=\! \sum_{\bf q} \hbar\omega_{\bf q}\,b^\dag_{\bf q} b_{\bf q} 
\!+\! \sum_{{\bf q}, \nu} \hbar n_{\nu} \big( g_{\bf q} b_{\bf q} \!+\! g^{\ast}_{\bf q} b^\dag_{\bf q}
                          \big) |\nu \rangle\langle \nu|.
\end{align}
The operator $b^\dag_{\bf q}$ ($b_{\bf q}$) creates (annihilates) a LA phonon
with wave vector ${\bf q}$ and energy $\hbar\omega_{\bf q}=\hbar c_s q$, 
where $c_s = 5110 \, \rm{m/s}$ is the longitudinal sound velocity.
$n_{\nu}$ denotes the number of excitons present in the state $|\nu\rangle$ and
$g_{\bf q}=\big(g_{\bf q}^{\text{e}}-g_{\bf q}^{\text{h}}\big)$
represents the exciton-phonon coupling constant that is given as the
difference between the electron-phonon and hole-phonon coupling constants 
\begin{align}
 g_{\bf q}^{\text{e(h)}}=\Psi^{e(h)}({\bf{q}})D_{e(h)}
                         \sqrt{\frac{ q}{2V\rho \hbar c_s}},
\end{align}
where $V$ is the sample volume, $D_{e}=-14.6 \, \rm{eV}$ and  $D_{h}=-4.8 \, \rm{eV}$
denote the deformation potential coupling constants of electrons
and holes and $\rho=5370 \, \rm{kg/m^3}$ is the material density. 
$\Psi^{e(h)}$ represents the
form factor of the carriers confined in the QD. For simplicity,
we assume a spherical harmonic oscillator confinement, leading to  
\begin{align}
 \Psi^{e(h)}({\bf{q}})=\exp(-q^2a^2_{e(h)}/4).
\end{align}
We choose an electron confinement length of $a_e = 3 \, \rm{nm}$ and set $a_h = 0.87 a_e$.

To analyze the combined carrier-phonon dynamics, we use a numerically
exact real-time path-integral approach, that accounts fully for all
non-Markovian effects and arbitrary multi-phonon processes. This
allows us to study the phonon-influence on the driven dynamics
at arbitrary temperatures and in parameter ranges, where approximate
methods come to their limits \cite{glaessl:11b}. A detailed
description of the formalism is given in Ref.~\onlinecite{vagov:11}.
Additional comments on the challenges that are faced within the
path-integral approach when accounting for more than two electronic
levels can be found in Ref.~\onlinecite{glaessl:12}.


\section{RESULTS}
\label{results}

\begin{figure*}[ttt]
 \includegraphics[width=17.cm]{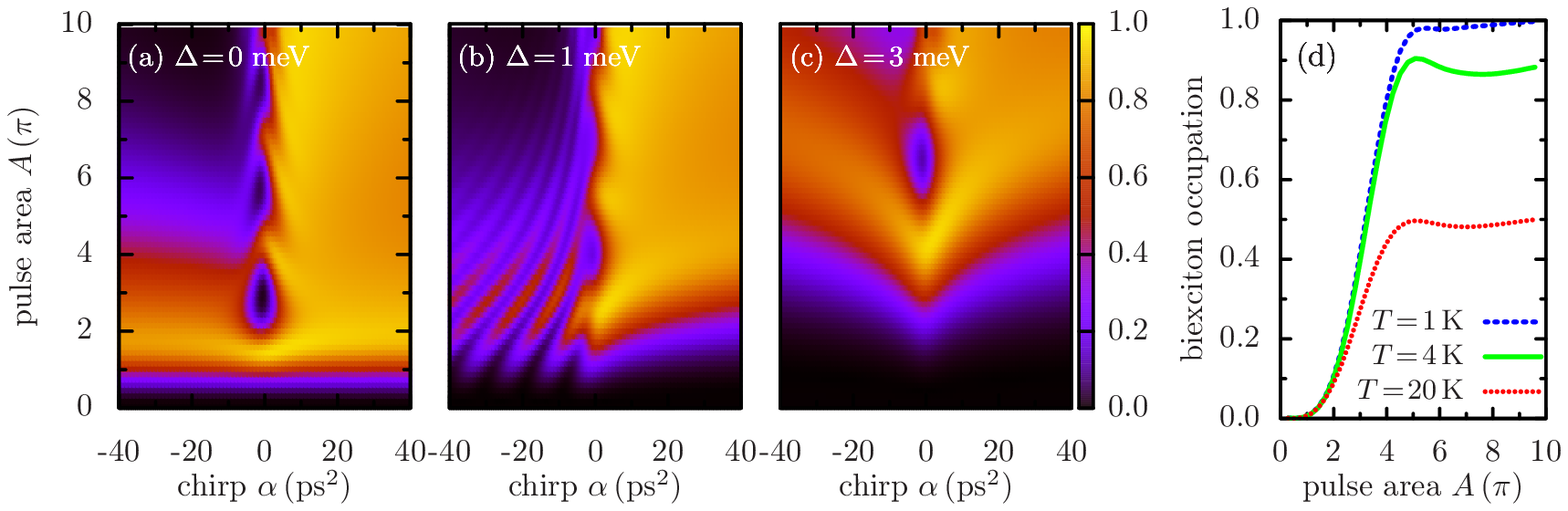}
    \caption{
             (Color online) Final biexciton occupation after a 
             linearly polarized and frequency-swept Gaussian pulse with $\tau_0=2\,\rm{ps}$
             within the two-photon resonance ARP-scheme.
             In (a)-(c) the occupation of the biexciton is shown as a 
             function of the original pulse area $A$ and the chirp $\alpha$
             for biexciton binding energies of (a) $\Delta = 0$, (b) $1$, and
             (c) $3\,\rm{meV}$ at $T=4\,\rm{K}$. In (d) the final occupation
             is plotted as a function of the original pulse area $A$ for
             $\Delta = 2 \, \rm{meV}$ and a chirp of $\alpha = 20\,\rm{ps}^2$
             at different temperatures.
            }
 \label{fig3}
\end{figure*}

In this section, we study how the presence of acoustic phonons affects
the preparation of the biexciton state by means of adiabatic rapid passage.
We shall analyze two protocols where in both cases 
 we will apply chirps ranging from $\alpha=-40$
to $+40 \, \rm{ps}^2$ and choose an original pulse length of $\tau_0=2\,\rm{ps}$
in order to use similar parameters as in recent experiments that reported
on a robust quantum dot exciton generation via ARP \cite{simon:11, wu:11}. 


\subsection{Two-photon resonance ARP-scheme}
\label{results:GBT}

Let us first concentrate on a two-photon resonance ARP-scheme that uses
one linearly polarized pulse. The central frequency of this pulse is chosen
such that for zero chirp the two-photon process is resonant to the ground
state biexciton transition, i.e., $\omega = \omega_0 - \Delta/(2\hbar)$, as
schematically sketched in the inset of Fig.~\ref{fig1}(a).

To provide a reference for the phonon influence on the efficiency of
this protocol, Fig.~\ref{fig1} shows the final biexciton occupation after the
pulse as a function of the original pulse area $A$ and the chirp coefficient
$\alpha$ for different values of the biexciton binding energy $\Delta$ neglecting
the exciton-phonon coupling. For zero chirp, the biexciton occupation depends
sensitively on the pulse intensity and performs Rabi rotations between the
ground and the biexciton state with a Rabi period that strongly depends on
the biexciton binding energy \cite{machnikowski:08}. 
In contrast, for large enough positive chirps, an ideal and robust biexciton
preparation, which is insensitive to small variations of the pulse area,
can be achieved provided that the applied pulse area exceeds the ARP-threshold.
The latter rises significantly with increasing $\Delta$ and shows for 
finite $\Delta$ also a considerable dependence on the strength of the chirp.
Interestingly, the situation is different for negative chirps: for $\alpha < 0$,
a stable biexciton preparation can be only achieved for almost vanishing or large enough
biexciton binding energies [cf. Figs.~\ref{fig1}(a) and (c)], but fails
for moderate values of $\Delta$, as it can be exemplarily seen from Fig.~\ref{fig1}(b)
for $\Delta=1$~meV, where for $\alpha < 0$ a stripe-like pattern forms. 
This dependence of the dynamics on the sign of the chirp contrasts to the
widely studied ARP-protocols for the generation of the exciton state
\cite{simon:11, wu:11, lueker:12, schmidgall:10}
and has to be taken into account also in situations where one expects
the influence of phonons to be negligible.

The results shown in Fig.~\ref{fig1} as well as the phonon impact
that will be studied below can be most easily
understood by analyzing the system evolution in the dressed state picture,
i.e., by considering the instantaneous eigenstates and eigenenergies that
are obtained by diagonalizing the combined light-matter
Hamiltonian $H_{\rm{dot,las}}^{\rm{lin}}$. The corresponding eigenenergies 
are plotted in Fig.~\ref{fig2} as a function of time for $\alpha=40\,\rm{ps}^2$
and different values of the original pulse area $A$ as well as of the biexciton
binding energy $\Delta$ together with the corresponding pulse envelope functions
(shaded areas). Long before and long after the pulse maximum, it is possible to
identify the instantaneous eigenstates with the electronic ground state $|G\rangle$,
the single exciton state $|X\rangle$, and the biexciton state $|B\rangle$, as it is
indicated in Fig.~\ref{fig2}. During the pulse, the character of the eigenstates
changes. For example, the lowest branch, that can initially be identified with
$|G\rangle$, eventually transforms into $|B\rangle$. It is precisely this 
transformation that allows a stable biexciton preparation within the two-photon
resonance ARP-scheme for positive values of $\alpha$ when the condition
for adiabatic passage is fulfilled, i.e., when the change in the frequency
and the change in the amplitude of the pulse are slow compared to the Rabi
frequency \cite{malinovsky:01}. In this case the system evolves along
a single branch and passes all anticrossings. Obviously, both conditions
are better fulfilled for higher pulse areas (corresponding to higher Rabi
frequencies), which can be nicely seen from Fig.~\ref{fig2} by comparing
the left and the right column, that show the branches for $A=2\pi$ and
$A=6\pi$, respectively: the splitting of the anticrossings considerably 
increases with rising $A$. By comparing the different rows of Fig.~\ref{fig2}
it can also be seen that for
a given pulse area, the anticrossings become narrower for larger biexciton
binding energies, explaining the rising threshold for ARP with increasing
$\Delta$ as shown in Fig.~\ref{fig1}. While for $A = 2\pi$ and $\Delta = 0\,\rm{meV}$ 
[cf. Fig.~\ref{fig2}(a)] the splitting of the central anticrossing at $t=0$ is 
large enough to ensure an adiabatic passage resulting in the preparation of the biexciton state
[cf. Fig.~\ref{fig1}(a)], the situation is different for larger $\Delta$. When
the anticrossing becomes too narrow [cf. Fig.~\ref{fig2}(c) and (e)], the
chirp rate is no longer small compared to the spacing between the branches,
the condition for ARP is violated and the system does not evolve along a single
branch.

Considering negative chirps, the situation is slightly more complex.
For $\alpha=-40\,\rm{ps}^2$ the system evolution along
the instantaneous eigenstates can easily be derived from Fig.~\ref{fig2} by 
reading the plots from the right to the left (i.e., by replacing the time $t$
by $-t$). Obviously, a transformation from $|G\rangle$ into $|B\rangle$ is
here possible by following the uppermost branch. In Figs.~\ref{fig2}(a) and (b)
for $\Delta=0$, one passes two anticrossings along the upper branch and the
system evolves completely adiabatically. For $\Delta=3\,\rm{meV}$,
both anticrossings are very narrow and the system jumps
two times from one branch to another branch instead of following the anticrossings.
Both crossings together result again in the preparation of the biexciton, as it
can also be seen in Fig.~\ref{fig1}(c). For $\Delta=1\,\rm{meV}$, the
anticrossings are narrow, but still considerably wider than for larger
biexciton binding energies. In consequence, the system does not evolve
completely adiabatically (via pure anticrossings) nor purely non-adiabatically
(via pure crossings) and a Rabi-like behavior
is realized instead of a stable and robust preparation of the biexciton 
giving rise to the stripe-like pattern as seen in Fig.~\ref{fig1}(b) for
$\alpha <0$. The details of this pattern can be explained as follows:
The fringes which develop from the occupation maxima or minima of the usual
Rabi rotations that are realized at $\alpha=0$ decline for rising chirps
because of two reasons: first, with increasing chirp the effective pulse area
exceeds more and more the original pulse area [as it can be seen by comparing Eqs.~(\ref{eq:f0}) and
(\ref{eq:f})] and second, the Rabi period decreases with increasing detuning \cite{allen:75}.
This leads to an increased number of occupation maxima or minima compared to $\alpha=0$.
 The attenuation of the stripe pattern
in the range of high pulse areas and large chirps is due to an increasing
splitting between the adiabatic branches that leads to a more and more
adiabatic evolution.

Next, we shall include the carrier-phonon interaction. Shown in
Figs.~\ref{fig3}(a), (b), and (c) is the final biexciton occupation after
the pulse as a function of the original pulse area A and the chirp coefficient
$\alpha$ at a temperature of $T = 4$~K for the same biexciton binding
energies as considered in Fig.~\ref{fig1}. Compared to the ideal
evolution in the absence of acoustic phonons there are similarities as
well as obvious differences. As in the phonon-free case the ARP-threshold
in the pulse-area rises with increasing biexciton binding energies and
for $\Delta = 1$ meV and negative chirps, the stripe-like pattern discussed above
is still visible. However, it is clearly seen that within the full model,
the efficiency of the protocol does now strongly depend on the sign of the
chirp, regardless of the size of the biexciton binding energy. 
While for positive chirps a rather stable biexciton generation is reached,
negative chirps result in general in very low
occupations of the biexciton state, in particular at high pulse areas.
To understand this difference, one should recall that for a positive chirp,
the adiabatic evolution follows the lowest adiabatic branch (cf. Fig.~\ref{fig2}).
Phonon mediated transitions to one of the other two branches, that spoil
the purely adiabatic evolution and hamper the preparation of the biexciton
state, are for $\alpha>0$ thus only possible via phonon absorption processes. 
At $4$~K the latter are rather weak and hence, there is a perceivable but small deterioration 
of the protocol efficiency compared to the phonon-free case. For negative 
chirps, in contrast, the adiabatic evolution of the system follows the 
uppermost branch and transitions to one of the lower branches are 
possible by phonon emission processes, which at low temperatures 
dominate. In consequence, the population of the upper branch is  
drastically reduced in the course of time and an efficient preparation of
the biexciton via ARP is no longer possible.

The temperature-dependence of the efficiency of the two-photon ARP scheme
is illustrated in Fig.~\ref{fig3}(d), where the final biexciton occupation is 
shown as a function of $A$ for a chirped pulse with $\alpha = 20\,\rm{ps}^2$, 
a biexciton binding energy of $\Delta = 2$~meV and three different temperatures
of $T=1$, $4$ and $20$~K. While for $4$~K and pulse areas above the ARP-threshold
occupations of about $90 \%$ are reached, the situation drastically
deteriorates at elevated temperatures, where phonon absorption processes gain in
importance due to a higher number of thermal phonons. Already at $20$~K, the
biexciton generation turns out to be very inefficient with typical 
biexciton occupations staying below $0.5$. On the contrary, it is interesting
to note that for temperatures well below $4$~K, an almost ideal biexciton
preparation can be realized as exemplarily shown for $T = 1$~K, where
in the adiabatic regime a high-fidelity preparation of the biexciton with
occupations higher than $0.98$ is achieved. This remarkably strong 
difference between $1$ and $4$~K contrasts to wide-spread expectations
that for low temperatures the impact of phonons is not very sensitive
to a variation of few Kelvins and emphasizes the importance of realizing
low temperatures. 

A closer look on the curves plotted in Fig.~\ref{fig3}(d) reveals further,
that for pulse intensities above the threshold, the biexciton occupation
depends nonmonotonically on the pulse area: an initial decrease is followed
by a slight increase. This nonmonotonic behavior in the ARP regime corresponds to the
reappearance of Rabi rotations in the usual Rabi regime \cite{vagov:07,glaessl:11}
and can be traced back to the resonance character of the carrier-phonon
coupling \cite{machnikowski:04}. The slow increase of the signal at high
pulse areas as shown in Fig.~\ref{fig3}(d) is mainly due to the Gaussian
pulse envelope, which in the Rabi regime is known to lead to a
rather weak reappearance because of an incomplete dynamical decoupling
between lattice vibrations and electronic dynamics. \cite{glaessl:11}
%
%
  
\subsection{Two-color pulse ARP-scheme}
\label{results:2PP}


\begin{figure*}[ttt]
 \includegraphics[width=17.cm]{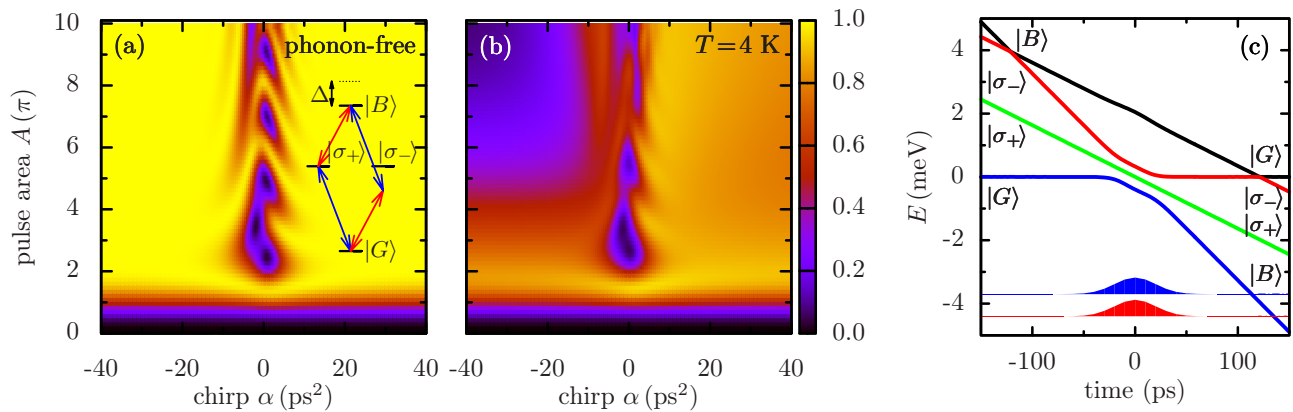}
    \caption{
             (Color online) Results for the two-color pulse ARP-scheme 
             that is schematically sketched in the inset of panel (a).
             In (a) and (b) the final biexciton occupation is shown as a 
             function of the original pulse area $A$ of each circularly polarized pulse and the chirp $\alpha$
             for $\tau_0=2\,\rm{ps}$ and a biexciton binding energy of $\Delta = 2\,\rm{meV}$
             in the phonon-free case (a) and in the full model at $T=4\,\rm{K}$ (b).
             Shown in (c) is the evolution of the adiabatic spectral branches 
             for $\tau_0 = 2 \,\rm{ps}$, $A=4\pi$, $\alpha = 40 \, \rm{ps}^2$ and $\Delta = 2\,\rm{meV}$.
             The evolution for $\alpha = -40 \, \rm{ps}^2$ can be read off the plot by following
             the branches from the right to the left. The shaded areas represent
             the envelope functions of both pulses, that have been shifted downwards for clarity.
            }
 \label{fig4}
\end{figure*}

Let us next turn to an alternative ARP-scheme. The QD is now driven by two
simultaneously applied frequency swept pulses of equal chirp, where the first pulse is circularly
$\sigma_+$ polarized and at the pulse maximum resonant to the ground state exciton transition,
i.e., $\omega_1 = \omega_0$, while the second pulse is circularly $\sigma_-$
polarized and at its maximum in resonance to the exciton biexciton transition, i.e.,
$\omega_2 = \omega_0 - \Delta/\hbar$. A schematic sketch of this two-color
protocol is drawn as an inset in Fig.~\ref{fig4}(a).
Neglecting the exciton-phonon coupling, also the two-color scheme allows for an ideal
and stable preparation of the biexciton. As an example, Fig.~\ref{fig4}(a) shows
the final biexciton occupation as a function of $A$ and $\alpha$ for $\Delta=2$~meV,
where $A$ now denotes the original pulse area of each circularly polarized pulse. 
We checked that even considerable deviations between both pulses with regard to the pulse
intensity or the frequency sweep do not reduce the efficiency of the scheme provided 
that both components fulfill the condition for ARP. Also a finite delay between both 
pulses has no significant effect on the efficiency as long as the pulse that drives the
ground state exciton transition precedes the pulse driving the exciton biexciton 
transition. Thus, the protocol is robust with respect to these changes and therefore, it is justified
to concentrate on the case of equal pulse parameters and zero delay for reasons of simplicity.

An analysis of the adiabatic spectral branches for the two-color protocol shares 
many similarities with that given in Sec.~\ref{results:GBT} for the two-photon resonance scheme 
and shall therefore not be presented at length. For positive chirps, the ground state
transforms into the biexciton state via the lowest branch, whereas for negative
chirps, the same transformation can be achieved by following the uppermost branch
[cf. Fig.~\ref{fig4}(c)]. 
Therefore, in the presence of the carrier-phonon coupling, a stable preparation is again
only possible for positive chirps, as illustrated in Fig.~\ref{fig4}(b) for $T=4$~K.
A further similarity to the first protocol is that the two-color scheme is even in the 
phonon-free case sensitive to the sign of the chirp and that for negative chirps and small
biexciton binding energies, a Rabi-like behavior is realized due to a neither purely
adiabatic nor purely non-adiabatic evolution. The resulting pattern seen in the
occupation when recorded as a function of $A$ and $\alpha$ (not shown) resembles that discussed
in Fig.~\ref{fig1}(b). Here, these patterns are most pronounced for $\Delta \sim 0.5$~meV,
which is half of the value of $\Delta \sim 1$~meV, where the Rabi-like behavior is
strongest in the two-photon resonance scheme. For these values of $\Delta$, the 
detuning of the respective off-resonant levels is identical for both protocols.   

\begin{figure}[bbb]
 \includegraphics[width=8.5cm]{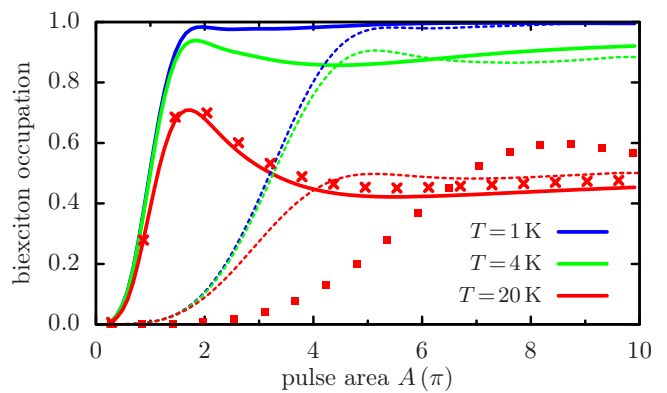}
    \caption{
             (Color online) Final biexciton occupation after  
             the two-color pulse ARP-scheme (solid lines) and the 
             two-photon resonance ARP-scheme (dashed lines) as a function of the
             original pulse area $A$ for $\Delta = 2 \, \rm{meV}$ and 
             $\alpha = 20\,\rm{ps}^2$ at different temperatures. 
             Crosses and squares represent results for $\Delta = 5 \, \rm{meV}$, $\alpha = 20\,\rm{ps}^2$
             and $T=20$~K for the two-color pulse ARP-protocol and the two-photon
             resonance ARP-protocol, respectively.
            }
 \label{fig5}
\end{figure}

Despite these similarities, one can also note significant and interesting differences
between both considered protocols. Unlike for the two-photon resonance scheme,
for the two-color protocol the pulse area threshold that must be exceeded
in order to ensure an adiabatic dynamics is independent of the biexciton binding
energy, as the transition to the biexciton does never involve excitation paths 
via strongly detuned intermediate levels. Independent of the chirp, the threshold
is roughly given by $A=2\pi$.
Therefore, the two-color protocol is particularly advantageous for QDs with large
biexciton binding energies. For example, for a high but still realistic value of
$\Delta=5$~meV and a chirp of $\alpha=20\,\rm{ps}^2$, the two-photon resonance
scheme would require pulse areas above $8\pi$ to reach the adiabatic limit
and thus sixteen times as intense pulses as within the two-color protocol.
This is exemplarily shown in Fig.~\ref{fig5} for $T=20$~K. 

Besides the threshold characteristics, the detailed temperature and pulse area
dependencies in the respective ARP-regime differ. This is shown in Fig.~\ref{fig5}, where the biexciton
occupation is plotted for both protocols as a function of the pulse area for a
chirp of $\alpha=20\,\rm{ps}^2$ and at temperatures of $1$, $4$, and $20$~K.
Solid lines represent the results within the two-color scheme and dashed 
lines the results of the two-photon resonance protocol that have already
been discussed in Fig.~\ref{fig3}(d). While for $T=20$~K the two-color scheme
is for pulse areas around $2\pi$ much more efficient than the two-photon
resonance protocol, the situation is reversed at high pulse areas, where
the efficiency is worse for the two-color scheme. For low
temperatures, the two-color protocol performs slightly better, in particular
at high pulse areas, where for $T=1$~K an almost perfect inversion to the
biexciton state is achieved. Common to the two-photon resonance scheme is the
pronounced difference in the protocol efficiency between $T=1$ and $4$~K as
well as the nonmonotonic pulse area dependence. However, for the two-color scheme,
the latter is slightly stronger and shows minima at smaller pulse areas.
This finding is in line with recent calculations for transform limited pulses,
that have revealed that the reappearance of Rabi rotations sets
in at smaller pulse areas, when the polarization of the excitation is 
changed from linear to circular.\cite{glaessl:12c}

%
%


\section{CONCLUSIONS}
\label{conclusions}
We have analyzed the impact of acoustic phonons on 
two protocols aiming at a robust preparation
of the biexciton state in a semiconductor QD.
In the absence of the carrier-phonon interaction 
either driving the system
with a single linearly polarized chirped pulse 
tuned at its maximum to the two-photon resonance
of the ground state biexciton transition
or using a two-color scheme where two circularly polarized chirped pulses
are at their maxima in resonance with the ground state exciton and the exciton biexciton transition,
respectively, enables a perfect preparation of the biexciton state
provided that the conditions for adiabatic rapid passage are fulfilled.
The most remarkable difference between both schemes, which is seen already without phonons, 
is the finding that the threshold for ARP rises drastically for the first protocol with
rising biexciton binding energies, while in the second protocol there is essentially no such
dependence on the threshold. Different from
related investigations of the preparation of the exciton state we here find
a strong asymmetry with respect to the sign of the frequency sweep already
in the idealized phonon-free case. In particular, we observe for finite negative
chirps and intermediate biexciton binding energies a Rabi-like behavior,
indicating a regime where the system neither evolves fully adiabatically
along the adiabatic spectral branches nor does it switch between different
branches without interference as it 
would  be the case in the extreme non-adiabatic limit.

Accounting for phonons in general reduces the fidelity of both protocols
and introduces an additional asymmetry with respect to the
sign of the chirp that, as also known from earlier studies of the exciton preparation,
results from the fact that for positive chirp the phonon emission is
suppressed. A robust biexciton preparation is thus restricted to
positive chirps and low temperatures. It is interesting to note that
already at 4 K the efficiency is reduced to about 90 \% while at
1 K an almost perfect biexciton preparation is possible in spite of
the phonons.

Taking the efficiency of the biexciton preparation as
the only criterion it turns out that the two-color scheme performs
slightly better for low temperatures. Moreover, for all temperatures
the threshold is lower. Nevertheless, evaluating both schemes
for practical usability these advantages should be weighted against
some obvious drawbacks: the need of two pulses with different colors
and (when thinking of using the biexciton decay cascade) 
the problem of laser stray light at the detection wavelength.
The latter problem can be overcome by using up-conversion techniques \cite{paillard:01}
or non-resonant emitter-cavity coupling \cite{ates:09}. Thus, by paying the price of
an increased effort one can indeed benefit from the better performance
of the two color scheme. We expect that our analysis shall be a helpful
guidance for making an adequate decision between performance and effort
in upcoming experiments.


\section{ACKNOWLEDGMENTS}

M.~G. is grateful for financial support from the Studienstiftung
des Deutschen Volkes. M.~D.~C. is supported by the Alexander
von Humboldt Foundation. P.~M. and T.~K.
gratefully acknowledge financial support in the framework of a Research
Group Linkage Project of the Alexander von Humboldt Foundation.


\end{document}